\documentclass[%
 reprint,
 amsmath,amssymb,
pra,
floatfix,
showkeys
]
{revtex4-2}

\usepackage{graphicx}
\usepackage{dcolumn}
\usepackage{bm}
\usepackage[breaklinks]{hyperref}
\usepackage{xcolor}
\hypersetup{
    colorlinks = true,
    citecolor={blue},
    linkcolor={red},
    allbordercolors = {white},
    urlcolor={blue}
}
\usepackage[position=top]{subfig}
\usepackage[justification=justified,format=plain,singlelinecheck=false,textfont=small]{caption}
\usepackage{booktabs}
\usepackage{makecell}
\usepackage{multirow}
\usepackage{siunitx}
\usepackage{ragged2e} 
\DeclareCaptionJustification{justified}{\justifying}

\begin{document}

\title{Reducing the number of single-photon detectors in quantum key distribution networks by time multiplexing}

\author{Jakob Kaltwasser}%
\affiliation{Institute for Applied Physics, Technische Universität Darmstadt,\\Schlossgartenstra\ss e 7,  64289 Darmstadt, Germany
}

\author{Joschka Seip}%
\affiliation{Institute for Applied Physics, Technische Universität Darmstadt,\\Schlossgartenstra\ss e 7,  64289 Darmstadt, Germany
}

\author{Erik Fitzke}%
\affiliation{Institute for Applied Physics, Technische Universität Darmstadt,\\Schlossgartenstra\ss e 7,  64289 Darmstadt, Germany
}

\author{Maximilian Tippmann}%
\affiliation{Institute for Applied Physics, Technische Universität Darmstadt,\\Schlossgartenstra\ss e 7,  64289 Darmstadt, Germany
}

\author{Thomas Walther}
\email{thomas.walther@physik.tu-darmstadt.de}
\affiliation{Institute for Applied Physics, Technische Universität Darmstadt,\\Schlossgartenstra\ss e 7,  64289 Darmstadt, Germany
}

\date{\today}

\begin{abstract}

We demonstrate a method to reduce the number of single-photon detectors (SPDs) required in multi-party quantum key distribution (QKD) networks by a factor of two by using detector time multiplexing (DTM).
We implement the DTM scheme for an entanglement-based time-bin protocol and compare QKD results with and without DTM in our QKD network with four users. When small efficiency losses are acceptable, DTM enables cost-effective, scalable implementations of multi-user QKD networks.

\end{abstract}


\maketitle

\section{Introduction}

Fundamental and technical advances in combination with existing quantum algorithms such as Shor's algorithm~\cite{shor_algorithm}, will enable quantum computers to break the current asymmetric encryption schemes~\cite{Grimes2019, Gerjuoy2005, Cheung2008}.
One promising way to restore security is to use quantum key distribution~(QKD) in conjunction with symmetric encryption methods~\cite{Gisin_2002, Scarani_2009, Xu_2020}.

In recent years, various QKD protocols, methods and networks have been demonstrated~\cite{Marcikic_2004, Chen_2021_2, Liao_2018, Yin_2017, Boaron_2018, Pittaluga_2021}. One main experimental challenge of practical QKD setups is to provide a high quantum key rate sufficient to encrypt the high data rates achieved in today's digital communication. 
Despite progress in the realization of long distance fiber-based~\cite{Chen_2021, Boaron_2018, Chen_2020, Wang_2022, Pittaluga_2021, Tang_2016} and satellite-based two-party QKD~\cite{Liao_2018, Chen_2021_2}, the distance between the parties still remains a major challenge due to the transmission losses as long as quantum repeaters are not accessible in a scalable manner~\cite{Pirandola_2017}. Of course, large distances in QKD can be achieved when resorting to networks with trusted nodes~\cite{Chen_2021_2}.

Another major challenge in the further development of practical QKD-systems is the scalability regarding the number of users, i.e.\ to ensure that a high number of communication parties can be connected. Only with well-developed QKD networks providing keys for many users, the technology will become relevant for a wide range of applications.
Each pair of parties exchanging a key should not have to trust the other parties connected to the network.
This can be achieved by implementing entanglement-based protocols in combination with wavelength-division multiplexing (WDM). 
To address both, the scalability in the number of users and the compatibility with existing telecom infrastructure, our group has recently demonstrated a robust, entanglement-based multi-user QKD network operating around \SI{1550}{\nano \meter}~\cite{Fitzke2022}. 

In the present paper, we provide a method to further improve the scalabilty of this network by halving the required number of single-photon detectors (SPDs). The SPDs are the major cost driver for such networks and reducing their number greatly reduces the cost for implementation. 
Therefore, our approach is based on detector time multiplexing (DTM) and allows to reduce the necessary number of SPDs per receiver unit from two to one.

\begin{figure}[bt]
    \includegraphics[width=0.9\linewidth]{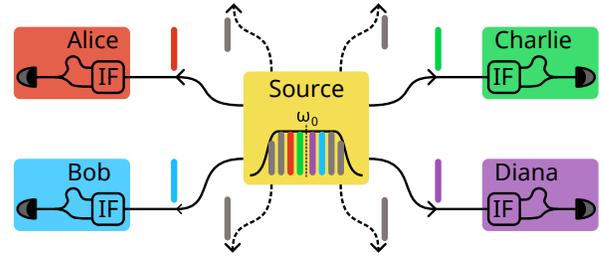}
    \caption{Operating principle of an entanglement-based star-shaped QKD network combining WDM with DTM. Multiple users are connected to the central photon pair source, where the entangled photons cover a broad spectrum and are demultiplexed into several frequency channels. Two users connected via channels with symmetric spacing around the central frequency $\omega_0$ receive entangled photon pairs due to energy conservation in the SPDC process. The receiver units are equipped with identical unbalanced interferometers (IFs). Due to DTM, each receiver unit only needs one SPD.}
    \label{fig:qkd_network_overview}
\end{figure}

In the following, we will introduce the concept of DTM and demonstrate its implementation in our QKD network together with WDM (cf.\ figure~\ref{fig:qkd_network_overview}).
Furthermore, we thoroughly evaluate the performance of DTM compared to the regular QKD setup and find only a small reduction in the quantum key rates. Moreover, we identify the causes for these losses and show that they are in principle remediable to a significant extent.

\section{Setup And Concept}

The principle architecture of our setup is a star-shaped QKD network in which a high number of user pairs can be connected to a central photon source, allowing multiple pairs of users to simultaneously and independently exchange secret quantum keys.
The source generates entangled photon pairs with a broad type-0 spontaneous parametric down-conversion (SPDC) spectrum~\cite{Tanzilli_2016, Tanzilli_2016_2}, which is split by WDM into various frequency channels. Each user receives the photons from one such channel.
The basic concept of this network has already been successfully demonstrated to work in a telecom environment by our group~\cite{Fitzke2022}.
All connected parties can exchange keys pairwise with each other - independently of all other parties. Since the photon pairs generated from SPDC are entangled in frequency due to energy conservation, the frequencies symmetric to the left and the right of the central frequency $\omega_0$ of the SPDC spectrum are entangled, as depicted in  figure~\ref{fig:qkd_network_overview}. Hence, by assigning quantum channels symmetrically positioned around the central frequency $\omega_0$ to a pair of users, these users obtain entangled photons from which they derive their quantum key. Any user pairing is possible by assigning those channels of the photon-pair spectrum to the party pairs willing to exchange secret keys. 

The employed BBM92 protocol~\cite{BBM92,Brendel_1999, Tittel_2000} uses the distribution of photon pairs entangled in time and phase.
This protocol is very well suited to implement scalable wavelength-multiplexed multi-party networks~\cite{Fitzke2022}. A major advantage over polarization-entangled protocols is its independence from polarization greatly enhancing the robustness of the transmission. Furthermore, unlike pure phase coding protocols~\cite{Gisin_2002}, we do not need active phase modulators.

\begin{figure}[t]
    \includegraphics[width=\linewidth]{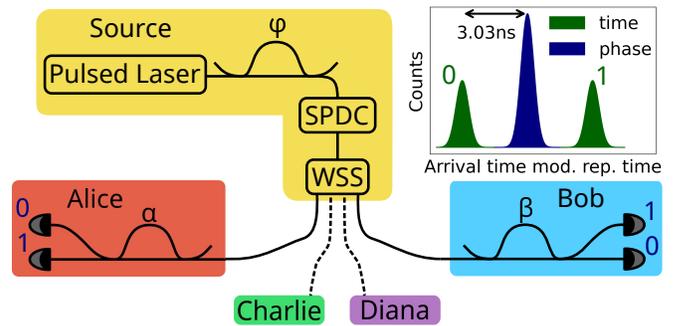}
    \caption{Schematic setup of our multi-party QKD with phase-time coding without DTM. The source (yellow box) and the exemplary receivers Alice, Bob, Charlie and Diana got identical unbalanced interferometers (only shown here for Alice and Bob), with a specified phase $\varphi \,, \alpha \,, \beta \,, \gamma \,, \delta$. We use a \SI{3,03}{\nano \second} delay in these interferometers. This delay generates a histogram of the photon arrival times, sketched at the top right used to gain key bits in the time basis, marked green. The phases of the interferometers are used to gain key-bits in the phase basis due to the entanglement of the photons. The phase basis is marked blue and the corresponding bits are tagged at the detectors.}
    \label{fig:ptc}
\end{figure}

The concept of phase-time coding for two parties is shown in detail in figure~\ref{fig:ptc}. As indicated, it can be easily adapted for more parties. We have implemented it for four parties. In this case the photon-pair spectrum is split up into four channels.

\begin{figure}[t]
    \subfloat[\label{fig:dtm_receiver_unit}]{
        \includegraphics[clip,width=\columnwidth]{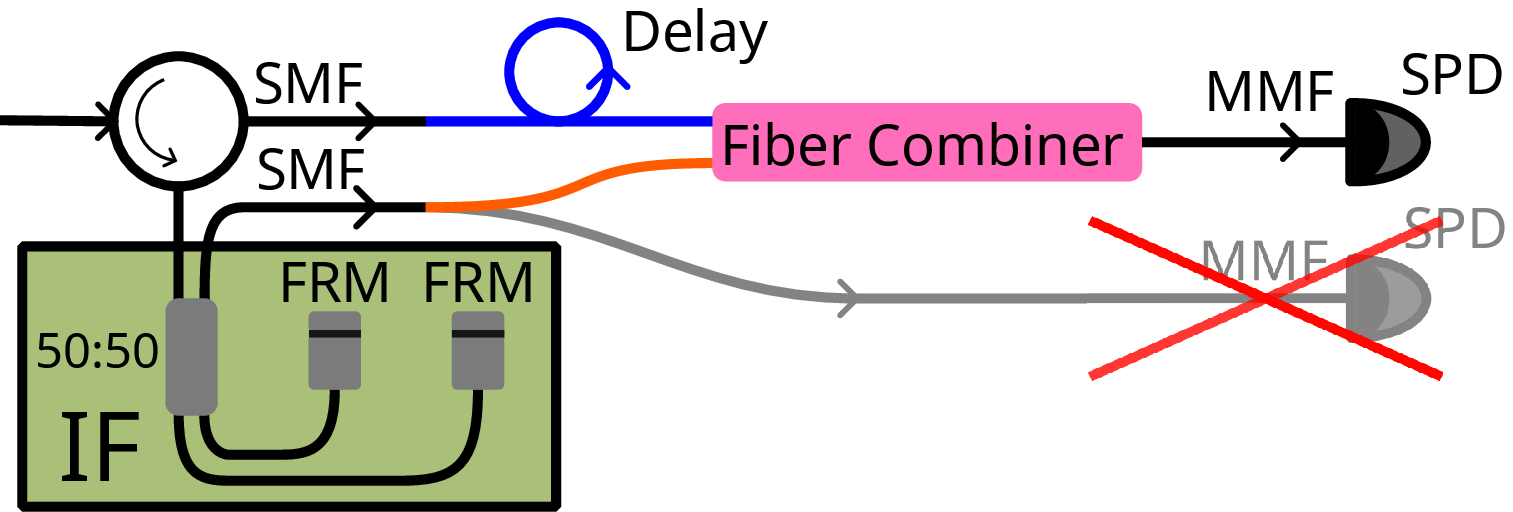}}\\
    \subfloat[\label{fig:dtm_peaks}]{
        \includegraphics[clip,width=\columnwidth]{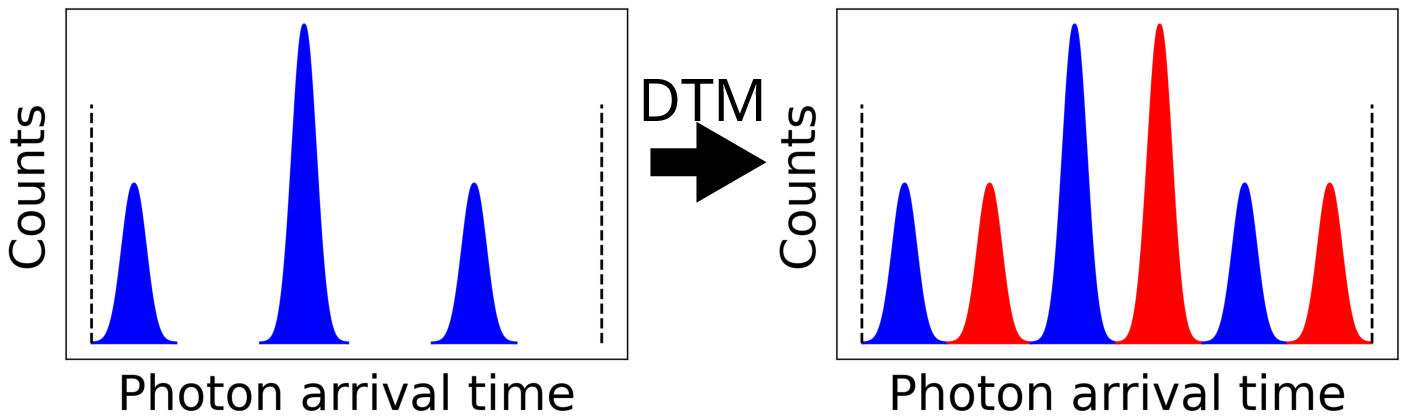}}
    \caption{(a) Scheme of a QKD receiver for phase-time coding. SMF - single-mode fibers with \SI{8.2}{\micro\meter} core diameter, MMF - multi-mode fibers with \SI{62.5}{\micro\meter} core-diameter. The temperature stabilized interferometer (IF) consists of a 50:50 beam splitter and two Faraday mirrors (FRM). In the usual setup, each IF output is connected to a separate SPD. For DTM, the receiver setup is modified: the interferometer outputs are combined by a 2x~SMF~$\rightarrow$~MMF fiber combiner.
    (b)~Resulting photon arrival histograms. Both interferometer outputs show the left histogram in the time domain. Due to the fiber section introducing a specific delay, one of the outputs is shifted in time, so that it fits into the free time intervals between the peaks of the other output (histogram in red) when combined with the other output by the fiber combiner.}
    \label{fig:dtm_all_changes}
\end{figure}

To implement the protocol, we use a photon-pair source containing a laser generating \SI{300}{\ps} long pulses at \SI{1550}{\nano \meter} with a  repetition frequency of \SI{109,89}{\mega \hertz}. These pulses then pass through an imbalanced interferometer, transforming them into well-separated double-pulses, with a specific phase relation.
These pulses are then frequency-doubled in a second harmonic generation stage.
Finally, we use SPDC~\cite{Tanzilli_2016, Tanzilli_2016_2} to generate the photon pairs in a fiber-coupled periodically-poled lithium niobate~(PPLN) crystal.
The photons are demultiplexed into the respective frequency channels and sent to a pair of parties, who want to exchange a secret key. 
For WDM, we use a wavelength-selective switch (WSS) allowing to arbitrarily swap the communication pairs.
In the users' receiver stations the photons each pass through another interferometer with an identical delay as in the source interferometer. Finally, the photons are detected in single-photon detectors (\emph{ID Quantique} ID220) connected to the two interferometer outputs.

In the time basis, three arrival times are possible per repetition cycle determined by the paths in the interferometers of the source and at the receiver: early arrival (short path in both interferometers), late arrival (long path in both interferometers) and arrival during the central time bin as a mixture of short and long paths, respectively.
A qualitative arrival-time histogram is shown in figure~\ref{fig:ptc}.
The key bits in the phase basis are given by the correlation between the two interferometer outputs $A_i$ at Alice's interferometer and $B_j$ at Bob's interferometer, with $i\,, j = 1 \,, 2$. Their detection probability in the central time bin depends on the sum of the phases of the source- and receiver interferometers~\cite{Marcikic_2004}:
\begin{align}
  P_{A_i,\,B_j}(\alpha,\beta,\phi) =  \frac{1}{4}\left(  1 + (-1)^{i+j}\cos \left( \alpha + \beta - \varphi \right)\right)
\end{align}

In the standard configuration, each receiver unit requires two SPDs. However, by employing DTM the necessary number of SPDs is cut in half.

\subsection{Detector Time Multiplexing}

Detector time multiplexing (DTM) has been recently used in setups to build photon-number resolved detectors~\cite{Achilles_2018, Fitch_2018, Natarajan_2013} and measuring higher-order photon correlations~\cite{Avenhaus_2010}. In these applications, the splitting of a pulse into several distinguishable time bins is used to obtain information about the photon number of the pulse via the detection probability per bin.

We use DTM to reduce the number of detectors per receiver module. Due to the relatively short (\SI{300}{\ps} long) laser pulses, the width of the peaks in the histogram is much less than the time delay in the interferometers of \SI{3,03}{\nano \second}. Therefore, there are unused time intervals in between the peaks, even when the pulses broaden in time due to chromatic dispersion in the transmission links.
Since the free time intervals themselves are longer than a peak width, further time bins can fit into the free intervals without interfering with the primary time bins. This can be used, for example, to increase the key rates by doubling the repetition rate of the photon source from \SI{109,89}{\mega \hertz} to \SI{219,78}{\mega \hertz}~\cite{Fitzke2022}. Alternatively, it can be used to realize DTM by combining both interferometer outputs into one fiber. In figure~\ref{fig:dtm_peaks} the peak structure of a single interferometer output is shown on the left and the final peak structure with DTM on the right. With DTM, it becomes a six-peak structure, where one three-peak interferometer output is nested into the other. The shift between the peak structures is achieved by introducing a fiber with a specific length producing a delay in one of the outputs before the interferometer outputs are combined.

One way to combine the outputs is to use a 50:50 tap coupler. But this will lead to additional losses of half of the photons, thus significantly decreasing the key rate.
Alternatively a fiber combiner can be used to combine two single-mode inputs into one multi-mode output.
We use this principle to join the two single-mode interferometer outputs in our receivers into one multimode fiber (MMF), which is then connected to an SPD.
Figure~\ref{fig:dtm_receiver_unit} displays the current setup of the receiver units and the necessary changes to implement DTM. The combiners we use are commercially available off-the-shelf components introducing additional insertion losses of about \SI{5}{\percent}. 

With these physical changes implemented, it is possible to distinguish between the two interferometer outputs by the different arrival times of the photons of each output. The time bins can be assigned to two virtual detectors, which from then on are used in the data evaluation as the two detectors in the regular setup without DTM. Additionally, the DTM requires a change of the data acquisition and evaluation software because it has to be able to distinguish the joint interferometer outputs.
The virtual detectors can be assigned to the correct interferometer outputs using a cross-correlation evaluation of the first few exchanged key bits, which are discarded afterwards.
With the cross-correlation the absolute temporal offset between the virtual detectors can be identified.
Together with the knowledge of which interferometer output has the longer propagation time to the combiner, the outputs can always be correctly assigned.

\section{Results}

To evaluate the performance of the QKD setup with DTM, a key exchange lasting over four hours was performed.
Keys were exchanged simultaneously between Alice and Bob with transmission distances from the source of \SI{26,9}{\kilo \meter} and \SI{50,4}{\kilo \meter} as well as between Charlie and Diana with transmission distances of \SI{9,6}{\kilo \meter}  and \SI{20,4}{\kilo \meter}, respectively.
The \SI{26,9}{\kilo \meter} link to Alice is a deployed dark fiber link provided by \emph{Deutsche Telekom} (cf.\ Ref.~\cite{Fitzke2022}). All other fibers are spooled fibers in the laboratory.

\begin{figure}
    \centering
    \includegraphics[width=\linewidth]{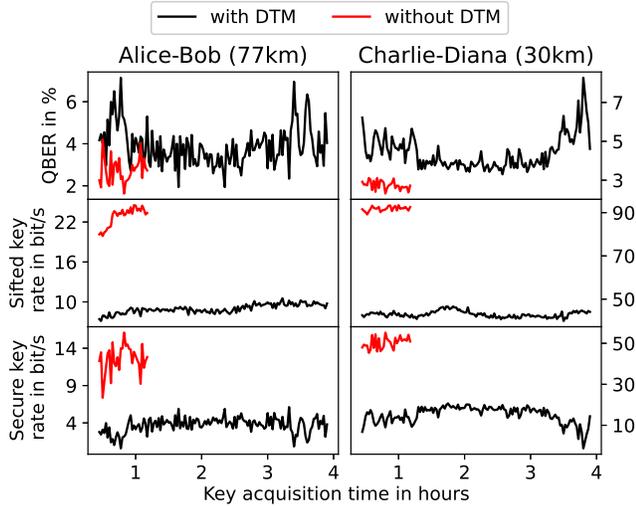}
    \caption{QKD results with and without DTM for our 4-party QKD setup.
    The data with DTM were acquired over an extended time period to demonstrate the long-term stability. 
    The key rates show a systematic difference between the measurements with DTM and without DTM. Also note the different y-scale for Alice~-~Bob and Charlie~-~Diana due to the different losses introduced by the different fiber lengths.
    }
    \label{fig:results_id220}
\end{figure}

The resulting quantum bit error rate (QBER) and sifted key rate are displayed in figure~\ref{fig:results_id220} and  average values are tabulated in table~\ref{tab:avg_rates}.
The secure key rate $l_\text{sec}$ was estimated from the sifted key rate and the QBER was estimated as described in Ref.~\cite{Fitzke2022} using a formula from Refs.~\cite{tuprints14042, Elkouss_2009}.

\begin{table}[ht]
    \caption{Overview over the average key rates and quantum bit error rates (QBERs) from figure~\ref{fig:results_id220} with the user pairs Alice-Bob and Charlie-Diana with and without DTM.}
    \label{tab:avg_rates}
    \centering
    \begin{tabular}{l@{\hskip 10pt}l@{\hskip 10pt}r@{\hskip 10pt}r@{\hskip 10pt}r}
    \toprule
         DTM & \makecell[l]{User\\combination}& \makecell[r]{Sifted\\key rate\\in bits/s} & \makecell[r]{QBER\\in \%} & \makecell[r]{Secure\\key rate\\in bits/s}\\
         \midrule
         \multirow{2}{*}{yes} & Alice - Bob & 8.1 & 3.89 & 3.6\\
         & Charlie - Diana & 43.0 & 4.37 & 15.3\\
         \midrule
         \multirow{2}{*}{no} & Alice - Bob & 22.8 & 2.74 & 12.5 \\
         & Charlie - Diana & 91.8 & 2.73 & 50.2\\
    \bottomrule
    \end{tabular}
\end{table}

The secure key rate with DTM is about \SI{70}{\percent} lower than without DTM. This is due to two effects: the major reduction in sifted key rate and the additional slightly increased QBER.

However, neither the lower sifted key rate nor the higher QBER could be attributed to the principle of DTM - at least not to this extent. 
The reason for the heavy drop in the sifted key rate and the large increase of the QBER was found in the dependence of our single-photon detectors' efficiencies on spatial modes in the MMF. 
The fiber combiner used in the DTM setup excites higher spatial modes in this fiber, to which our detectors are much less sensitive even though specified as multi-mode detectors.

This effect was verified with two MMFs spliced together with a small spatial misalignment of the cores to excite higher spatial modes. A simple power measurement was used to verify that
the insertion loss introduced by the splice is negligible.
This fiber is inserted into a setup where attenuated laser pulses were detected with the SPD. With the spliced MMF the count rates were around \SI{24}{\percent} lower compared to measurements with a regular MMF without such an offset splice due to the relatively large spatial mode dependency of the detection efficiency.

\begin{figure}[t]
    \centering
    \includegraphics[width=\linewidth]{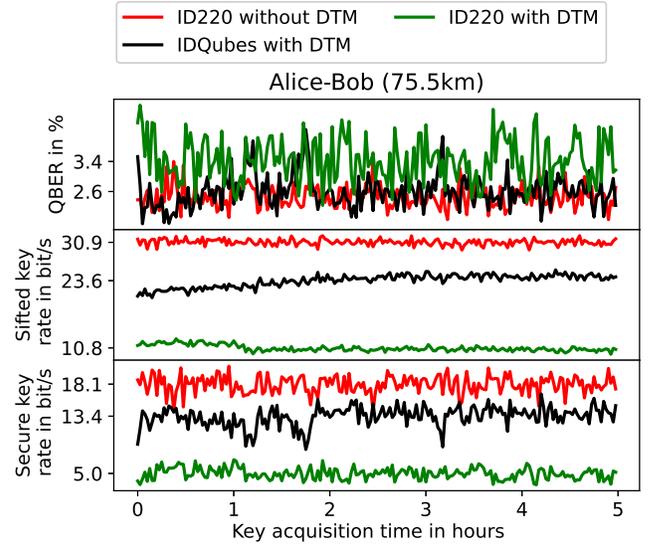}
    \caption{QKD results for a 2-party QKD setup with and without DTM. In the case of DTM, IDQubes are used additionally, since the ID220s were not able to detect different spatial modes equally efficient.
    }
    \label{fig:results_idqubes}
\end{figure}

After identifying this effect, a control experiment with QKD between two parties was performed with two IDQube detectors from \emph{ID Quantique} employing the same settings for detection efficiency and dead time as for the ID220 detectors. Figure~\ref{fig:results_idqubes} shows the results indicating that the effect does not occur with these detectors. The sifted key rates with DTM using IDQubes is more than twice as high as with the ID220s although
both detector types have approximately the same detection efficiency and dead time. This observation corroborates our assumption that the ID220s' lower detection efficiency for higher spatial modes significantly reduces the key rates in our DTM setup.

Nevertheless, using DTM still decreases the key rate by around \SI{23,6}{\percent} even when IDQubes are used. Two reasons were identified:
On the one hand DTM required additional fiber connections in our experimental setup introducing additional losses, and the fiber combiner leads to insertion losses of around \SI{5}{\percent} per receiver. This alone reduces the sifted key rate by around \SI{10}{\percent}. On the other hand DTM also leads to higher saturation of the detectors since now both interferometer outputs are detected at only one SPD. The effect of the saturation can be estimated using~\cite{Sat}

\begin{align}
    \frac{R_{\text{m}}}{R_{\text{e}}} \approx 1 - \tau \, R_{\text{m}} \,, \label{Sat_estimation}
\end{align}

with the measured count rate $R_{\text{m}}$, the expected count rate $R_{\text{e}}$ and the detector dead time $\tau$, which is $\SI{10}{\micro \second}$ in our setup.

Using ID220s without DTM, the detected count rates for Alice are around \SI[per-mode=symbol]{19500}{\per \second} and \SI[per-mode=symbol]{13500}{\per \second} at the two interferometer outputs. Bob receives count rates of around \SI[per-mode=symbol]{9500}{\per \second} and \SI[per-mode=symbol]{6600}{\per \second}. Using IDQubes with DTM, the detected count rate for Alice is around \SI[per-mode=symbol]{22000}{\per \second}. Bob receives a count rate of around \SI[per-mode=symbol]{12500}{\per \second}. This results in an additional decrease of the sifted key rate of around \SI{11}{\percent} due to saturation according to equation~\ref{Sat_estimation}.

The insertion loss of the additional components and the additional fiber connections amount to an efficiency loss of \SI{10}{\percent}. In combination, both effects give a decrease of efficiency of \SI{20}{\percent}, which essentially describes the measured difference of \SI{23.6}{\percent} between QKD with and without DTM. The remaining difference of \SI{3,6}{\percent} may be accounted to measurement uncertainties and additional fiber connections in the DTM setup and to the fact that equation \eqref{Sat_estimation} is only an approximation. Thus, DTM operates as expected. 

The QBER when using ID220 without DTM is similar to the QBER using IDQubes with DTM. However, the newer IDQubes also have lower dark count rates slightly reducing the QBER. Since we do not have access to four IDQubes, a check without DTM using IDQubes could not be performed and the exact effect of DTM on the QBER could not be determined. A possible cause for a slight increase of the QBER could be potential crosstalk between time bins due to the reduced gap between adjacent time bins. Additionally, with DTM crosstalk in the phase basis is also possible, because the time bins of the central peaks of both interferometer outputs lie directly next to each other in our case. It might be possible to reduce this effect by choosing a different time offset between the interferometer outputs so that the central peaks of the two outputs are further apart from each other, with a time-basis bin in between them.
In figure~\ref{fig:dtm_peaks} this would be the case, when shifting the red histogram, so that the left small peak of it is right to the blue big peak. This should reduce at least the crosstalk in the phase basis.

\section{Discussion and Outlook}
\label{Discussion}

Our experiments demonstrate the functionality of DTM, showing a good stability even for long key-exchange measurements over four to five hours.

However, we have observed a major decrease of sifted key rate in our first attempts. We attribute this effect to the sensitivity of the detection efficiencies of our SPD used on spatial modes.
A control measurement with a SPD, which detects all spatial modes equally efficient, has shown that in principle QKD with DTM works as expected and without unexpected losses in efficiency.
Consequently, a requirement for the DTM are SPDs which can efficiently detect different and in particular higher spatial modes.

Only a small increase in QBER in case of DTM is not fully explained yet and needs further investigation. For that a comparative measurement with and without DTM should be performed with the IDQube detectors.

A major effect reducing the sifted key rate is the saturation of the detectors, limiting the possible count rates for shorter distances between the parties. In this case, the photon rates arriving at the receiver units are higher. Naturally, since both interferometer outputs are fed into a single detector, DTM is not suitable for high count rates, since saturation becomes the dominant effect.
The saturation effect could be reduced by using other detector types with lower dead time, such that the ratio of the measured to the actual rate in formula~\ref{Sat_estimation} approaches unity and the sifted-key rate further approximates the case without DTM.
An exemplary alternative for other detectors would be commercially available superconducting-nanowire single-photon detectors (SNSPDs). These not only have a higher detection efficiency but also much lower dead times 
(e.g.\ IDQ ID281)~\cite{IDQ_SNSPD}. 

In terms of the number of detectors required for a larger network, our setup can keep up with measurement device independent (MDI) protocols which only require as many detectors as connected users~\cite{Chen_2020, Wang_2022}. At the same time, we maintain the advantage over MDI protocols in terms of photon source scalability: In our setup, a single central photon source serves all users. 

\section{Conclusion}

To the best of our knowledge, we have for the first time presented the concept of DTM as a solution to reduce the number of SPDs needed to one per receiver in a multi-party QKD-network. This reduction will significantly reduce the cost per receiver module and thereby increase the scalability of such networks in terms of connected parties. This is an important step towards affordable and practical QKD networks, and thus the possibility of widespread use of this technology addressing the growing threats of quantum computing on current encryption techniques. 

From a technical point of view, our setup is already highly scalable with regard to simultaneously connected parties due to a broad photon-pair spectrum in conjunction with WDM~\cite{Fitzke2022}. WDM in combination with entanglement-based protocols solves the challenge of serving a large number of parties through only a single source, while DTM drastically reduces the implementation cost of the network.

The experimental implementation of DTM was realized by inserting a fiber introducing a specific time delay after one of the interferometer outputs and combining both interferometer outputs with a fiber combiner.
The problems arising from spatial dependencies of our single photon detectors were identified and we have shown that the corresponding problems do not occur with detectors that properly detect all spatial modes with the same efficiency.

When considering the usage of SPDs which are able to detect higher spatial modes equally efficiently, there remains only one drawback for the DTM setup:
detector saturation plays a major role because omitting one detector leads to an increase of the photon rate for the remaining detector.
In our setup, DTM is therefore only useful at larger distances between the parties where the count rates are lower.

\begin{acknowledgments}

This research has been funded by the Deutsche Forschungsgemeinschaft (DFG, German Research Foundation), under Grant No. SFB 1119--236615297. We thank Paul Wagner from Deutsche Telekom Technik GmbH for lending us the WSS and fiber spools and Felix Wissel from Deutsche Telekom Technik GmbH for the provision of a dark fiber test link.

\end{acknowledgments}

\providecommand{\noopsort}[1]{}\providecommand{\singleletter}[1]{#1}%

\end{document}